.

# Chiral Dynamics of the Polarizing Fracture Functions for Baryon Production

Dennis Sivers

Portland Physics Institute
4730 SW Macadam #101
Portland, OR 97239

Spin Physics Center
University of Michigan
Ann Arbor, MI 48109

## Abstract

The concept of spin-directed momentum provides a useful and restrictive framework for describing dynamical mechanisms that can lead to single-spin observables. The value of this framework can be demonstrated by consideration of the polarizing fracture functions, $\Delta^N M^q_{B\uparrow/p}(x,z,k_{TN};Q^2)$, that characterize the production of polarized baryons in the target fragmentation region of semi-inclusive deep inelastic scattering from an unpolarized target. When Bjorken x is chosen large enough to indicate a hard scattering from a valence quark, the fracture function formalism dynamically selects a quark-diquark basis for baryon structure. Attention to constituent orbital angular momentum in the formation process and its role in contributing to the transverse momentum of the produced baryon illustrates important aspects of the generation of polarization observables.

## I. Introduction

Single-Spin observables can be used to study hadronic dynamics in many different ways. [1,2] A useful concept in this study program involves the identification of a spin-directed momentum in measurements involving a single-spin observable. Each and every single-spin measurement in a scattering process describes a momentum transfer whose direction is determined by the orientation of the measured spin. For parity-violating spin observables, that direction involves a helicity projection since the couplings of the W, Z bosons to flavor doublets of fundamental fermions select specific helicity states. Since helicity is a pseudoscalar observable, parity violating asymmetries lead to spin density matrices that can be diagonalized in a helicity basis. The value of this selection can be appreciated, for example, in the phenomenological analysis of spin asymmetries in W, Z production processes.[3]

The dynamical mechanisms leading to parity-conserving, transverse, spin observables are also easy to classify in this manner.[4] The spin-directed momentum specified by these observables can always be written in the form, $k_{TN} = \vec{k}_T \cdot (\hat{\sigma} \times \hat{p})$, where $\hat{\sigma}$ is a unit vector denoting the direction of the measured spin. Mulders and Tangerman [5] have identified four types of $k_T$-dependent hadronic operators that involve projections of this form leading to transverse single-spin asymmetries at leading twist in processes involving large momentum transfers. To perform calculations it is helpful to know that transverse single-spin asymmetries are odd under an artificially constructed symmetry, $A_\tau$, that is sometimes designated "naïve time reversal."[6,7] Using this symmetry it is then possible to define projection operators to demonstrate that transverse single-spin observables lead to spin density matrices diagonalizable in the transversity basis[8] just as parity-violating observables are diagonalizable in a helicity basis.

As indicated in Table 1, the quartet of $A_\tau$-odd functions identified by Mulders and Tangerman includes two types of $k_T$-dependent distribution functions. The orbital distributions for quarks, antiquarks and gluons describe the orbital motion of color constituents in a polarized nucleon while the Boer-Mulders functions[9] describe the correlations in an unpolarized ensemble of nucleons between the transverse polarization of light quarks and their internal orbital angular momentum. The classification also includes two types of fragmentation functions; the Collins functions[10] characterize the effect of orbital angular momentum on the fragmentation of transversely polarized light quarks into different hadrons while the polarizing fragmentation functions give the correlation between polarization and momentum for hadrons arising from the fragmentation of an initially unpolarized ensemble of constituents.

The Collins functions and the Boer-Mulders functions of Table 1, in addition to being odd under $A_\tau$, are also designated as <u>chiral-odd</u> functions because they involve the transverse polarization of light (u,d) quarks.[11] Since perturbative QCD processes involving light quarks preserve helicity, chiral-odd functions describing physical observables occur in the hard-scattering expansion of QCD only in pairs. Observables involving the Collins functions can, for example, appear in conjunction with the chiral-odd transversity distributions [12] and, hence, the determination of Collins functions can play an important role in understanding this important component of the transverse spin structure of the nucleon. Similarly, the Boer-Mulders functions must be observed in combination with other chiral-odd functions. In contrast, the orbital distributions and the polarizing fragmentation functions involve hadronic spins and can be measured in combination with unpolarized distributions. They are therefore designated <u>chiral-even</u>, or hadronic functions.

While the distinction between chiral-odd and chiral-even functions plays an important role in defining observables, the term "chiral dynamics" in this paper has a broader context. As we shall argue, all four sets of functions in the Mulders-Tangerman classification share a common dynamical origin. They all involve the orbital angular momentum of light quarks arising from the interplay of confinement and chiral dynamics in the nonperturbative regime of QCD. Since hadronic spins and quark spins are correlated, a systematic, quantitative description of any set of these functions offers the prospect of a detailed understanding of these important nonperturbative mechanisms. Spin-orbit effects involving color constituents thus provide a distinctive signal of chiral dynamics in QCD.

It turns out that the two fragmentation functions of the Mulders-Tangerman quartet provide more direct access to the underlying dynamical mechanisms than do the two distribution functions since, in the Collins functions and polarizing fragmentation functions, the orbital angular momentum generated by the chiral dynamics appears directly in the final state while in the Boer-Mulders distributions and orbital distributions of a stable nucleon, the orbital angular momentum is associated with virtual processes and the spin-directed momentum in a given measurement involving these distributions can be altered or hidden by initial-state and final-state interactions. The parameterization of the Collins functions for pions introduced by Artru, Czyzewski and Yabuki (ACY) [13] in the framework of the Lund string model [14] provides a striking example of the direct connection between orbital angular momentum and a spin-directed momentum in fragmentation. The parameterization can readily be extended to the production of other pseudoscalar mesons and, when applied to vector meson fragmentation functions, the approach dramatically illustrates the close connections existing between the Collins functions and the polarizing fragmentation functions.

It is interesting to apply this investigation to a variety of other processes. In this work, therefore, we therefore consider further extending the underlying dynamic approach described by Artru, Czyzewski, and Yabuki and apply it to the production of polarized baryons in the target fragmentation region of processes involving a hard scattering. Such semi-inclusive production processes can be described by the so-called fracture-functions introduced by Trentadue and Veneziano. [15,16] In particular, we are going to examine the set of polarizing fracture functions, $\Delta^N M_{B\uparrow/p}^{q_i}(x,z,k_{TN};Q^2)$ and $\Delta^N M_{B\uparrow/n}^{q_i}(x,z,k_{TN};Q^2)$ that characterize the spin dependence of polarized baryon production in semi-inclusive deep inelastic processes where the hard scattering involves a valence quark, $q_i = u,d$, from the target proton or neutron. This supplements the set of processes involving polarizing fragmentation functions to include a large set of observables with a rich, informative, structure. By choosing Bjorken x large enough to involve a valence quark in the hard scattering, the fracture function formalism dynamically selects the quark-diquark basis to describe the quantum numbers of the produced baryons. For clarity and economy, in this paper, we shall restrict attention to the production of S-wave baryons that share the diquark fragment of the target nucleon and contain only one additional u, d, or s quark generated in the fragmentation process. The set of produced baryons thus considered consists of the proton, neutron, Lambda and Sigma in the J=1/2 octet and the Delta and Sigma states of the J=3/2 baryon decuplet. Although the "fragmentation" of the remnant diquark system into a new baryon as described by the fracture functions contains a large amount of additional dynamical information, we focus primarily in this note on the role of orbital angular momentum in the baryon production mechanism and its ability to provide a transverse momentum separation between the baryon states produced with different polarizations. This simple, direct approach to polarized baryon production illustrates the role of single-spin observables in isolating specific mechanisms in chiral dynamics.

The remainder of the paper is structured as follows: Sec. II reviews the application of symmetries in the discussion of single-spin observables and how they can be used to determine an appropriate basis for the diagonalization of spin density matrices. Sec. III discusses the fracture function formalism and its application as a surrogate approach to diquark fragmentation. We introduce the generalization of the formalism to include spin and spin-directed momentum and describe the role of orbital angular momentum in the productions processes. Sec. IV gives the explicit expressions polarizing fracture functions in the ACY model for the many different states using spin and isospin Clebsch Gordan coefficients. Sec. V concludes with a brief discussion of the chiral elements in the underlying model calculations and includes a discussion on the possibility of incorporating diquark degrees of freedom into chiral models for the orbital distributions and Boer-Mulders functions of the nucleon.

## II.   Spin-Directed Momentum and the Classification Theorem.

Single-Spin observables are highly constrained by rotational invariance and finite symmetries. If we consider a scattering event in the x-z plane with initial state CM momenta along the z-axis the constraints on an initial state spin observable generated by finite symmetries are shown in Table 2. This table augments the usual finite symmetries of quantum field theory; C, P, T, by including a set of space-time operations that also involve the Hodge dual operator, *, for differential forms in 3+1 dimensions. The purpose of the augmentation is to allow the construction of a transformation, O, under which every single-spin observable is odd. This operator can then be used to form projection operators for spin calculations. The Hodge dual operator in differential geometry has the convenient property that it interchanges the role of 3-vectors and axial 3-vectors.

$$*:(V_i, A_k) \rightarrow (\tilde{A}_i, \tilde{V}_k) \qquad (2.1)$$

where $V_i$ denotes a set of 3-vectors and $A_k$ a set of axial 3-vectors and $\tilde{A}_i, \tilde{V}_k$ are the corresponding images in the dual space. [17]   Since the parity operator, $P$, changes the sign of 3-vectors while leaving axial 3-vectors unchanged,

$$P:(V_i, A_k) \rightarrow (-V_i, A_k) \qquad (2.2)$$

the action of the parity operator in the Hodge dual space, $\tilde{P} = P*$, is given by

$$\tilde{P}:(\tilde{A}_i, \tilde{V}_k) \rightarrow (\tilde{A}_i, -\tilde{V}_k) \qquad (2.3)$$

so that, transforming back to regular space gives

$$*\tilde{P}:(\tilde{A}_i, \tilde{V}_k) \rightarrow (V_i, -A_k) \qquad (2.4)$$

the result we desire. The spin-reflection operator $O = *\tilde{P} = *(P*)$ therefore has the property that on a 3+1 dimensional differential manifold it changes the sign of axial 3-vectors while leaving 3-vectors unchanged. This implies that O can be considered the complete snake operator. Just as a mirror gives a partial realization of the parity operator, the Siberian snake of accelerator physics invented by Derbenev and Kondratenko [18] provides a partial realization of this O- transformation in its action of rotating spins by $180^o$ while preserving the direction of momenta. The spin-reflection, or snake operator, O, as defined in (2.4) can then be combined with the parity operator in ordinary space-time to define the operator $A_\tau$, where $A_\tau = PO$. Combining (2.2) and (2.4) gives

$$A_\tau:(V_i, A_k) \rightarrow (-V_i, -A_k) \qquad (2.5)$$

Using these properties, it is possible to apply the operator $A_\tau$ to the scalar $k_{TN}$

$$A_\tau : k_{TN} = A_\tau : \vec{k}_T \cdot (\hat{\sigma} \times \hat{p}) = -k_{TN} \qquad (2.6)$$

to demonstrate the requirement that the operator $A_\tau$ selects the parity-conserving spin-directed momentum that identifies transverse single-spin asymmetries. It can be shown that the operators, $1, P, O$, and $A_\tau$ form an Abelian group with each operator being its own inverse. Since $O = PA_\tau$ and since each single spin observable is odd under $O$, we have the classification theorem.

*Classification Theorem for Single-Spin Observables*

Each single-spin observable falls into one of two distinct categories.

1. Odd under parity, $P$, and even under $A_\tau$.
2. Even under parity, $P$, and odd under $A_\tau$.

In the standard model, parity-odd spin observables involve point-like interactions with the W, Z bosons while $A_\tau$-odd observables involve fundamental fermion mass parameters or coherent, spin orbit dynamics. More discussion of the classification theorem can be found in the references [4,19].

The close connection between the scalar $k_{TN} = \vec{k}_T \cdot (\hat{\sigma} \times \hat{p})$ and spin-orbit effects is described in Figure 1. The quantity $k_{TN}$ is then seen to be even under P, C and T while being odd under the operator $A_\tau$. The other operators in Table 2, $O_\subset$ and $O_\supset$ can also be used to define projections when it is important to keep track of the fact that the dynamical mechanisms leading to transverse spin asymmetries preserve charge conjugation and time reflection. Since the processes we shall consider typically involve u, d quarks, the behavior under G parity, combining isospin and charge conjugation, is also useful. The idempotent projection operators

$$P_A^\pm = \frac{1 \pm A_\tau}{2} \qquad (2.7)$$

thus play an important role in practical calculations. Because of the result of Kane, Pumplin and Repko (KPR) [20] we know that the perturbatively-calculable, hard-scattering component of any calculation involving light (u,d) quarks contains no significant $A_\tau$-odd effects. This observation leads to what can be termed KPR factorization. The factorization property occurs because the spin-directed momentum defining the single-spin observable must necessarily be generated by a soft coherent process connected to the hadron containing the measured spin. From the discussion above, that coherent process involves an expectation value $\langle \hat{\sigma} \cdot \vec{L} \rangle$. Because the projection operators (2.7) apply both at the amplitude level and at the $|amplitude|^2$ level, the spin-density matrix generated by the $A_\tau$-odd dynamics is necesssarily diagonal in the transversity basis. We can see a direct application of these two important simplifications in the following discussion of the polarizing fracture functions for the production of baryons in the target fragmentation region.

### III.  Fracture Functions and the Role of Diquarks in Baryon Production.

   Fracture functions, introduced by Trendadue and Veneziano [15] in 1994, represent hybrid forms between structure functions and fragmentations functions. The fracture function, $M^b_{p/h}(x, z_h; Q^2)$, characterizes the conjoint probability for finding both the parton, b, with Bjorken $x = Q^2/2p.q$ and the hadron, h, with Feynman $z_h = p \cdot p_h / p \cdot q$ in a semi-inclusive deep inelastic scattering process from a proton target.  Depending on the kinematics and context, this probability density can be construed as describing an effective structure function for a virtual composite system with quantum numbers of $p\bar{h}$, or as characterizing the fragmentation of a target-remnant system with quantum numbers of $p\bar{b}$.[21]   We can include a range of single-spin observables into the fracture function formalism by considering a transverse spin measurement involving any of the external particles.  That means, for example, we can consider $M^{b\uparrow}_{h/p}, M^b_{h\uparrow/p}$, or $M^b_{h/p\uparrow}$.  As can be inferred from the discussion in Sec. I above, the fracture function $M^{q\uparrow}_{h/p}$ for a light quark is a chiral-odd object so that this combination must be construed as appearing in conjunction with another chiral-odd object.  This additional measurement can be left unspecified or it is possible to formulate a projection that essentially involves the convolution of a Boer-Mulders function and a Collins function without assuming a specific factorization. The intriguing versatility of this theoretical approach leads to a wide range of applications and a summary of the various techniques and of the calculations involving the fracture function formalism is well beyond the scope of this work.[22]

   The particular application that we will consider here selects events in which the hard-scattering process involves a valence quark from a proton or neutron.  This selection  requires a kinematic cut that selects the high-x region of the hard-scattering process but we will not, at this point, attempt to specify this cut in quantitative detail.  The specific purpose of the selection is to use the fracture function formalism to characterize a chromodynamic final state that initially contains valence quark quantum numbers in the current fragmentation region and diquark quantum numbers in the target fragmentation region.  In terms of SU(3) color dynamics, this set of dynamically generated $3_c, \bar{3}_c$ final states shares some characteristics of the $q, \bar{q}$ states produced in $e^+e^-$ annihilation and our goal is to use the fracture function as a surrogate for the fragmentation function of a diquark system into a baryon that approaches the underlying SU(3) color dynamics in a manner similar to the dynamics for the fragmentation of an antiquark into a meson. The two final states are, of course, not identical and to explore their similarities adequately we need to introduce some features of the chromodynamic description of diquarks.

Interest in the diquark structure for baryons has flourished because of solid phenomenological evidence. The venerated review by Anselmino, Predazzi, Ekelin, Fredricsson and Lichtenberg [23] provides a seminal set of important references. We will follow here a nomenclature inspired by the recent work of Jaffe and Wilczek.[24,25,26] The principle objects to be considered will be the $\bar{3}_c$ diquark systems

$$[q,q] \to (SU_3^{flavor} = \bar{3}, J^P = 0^+) \quad (3.1)$$
$$\{q,q\} \to (SU_3^{flavor} = 6, J^P = 1^+) \quad (3.2)$$

The notation is chosen to represent the symmetry of the diquark system in flavor space so that [ , ] denotes antisymmetric and {,} denotes symmetric. In the following we will use the familiar symbols for quark flavors when convenient. For example, [u,d], [d,s] and [s,u] describe the $\bar{3}^{flavor}$ states (note that the ordering is important because of the antisymmetry) while {u,u}, {u,d}, {d,s}, etc. are elements of the $6^{flavor}$ (here the ordering of the symbols denoting flavor does not matter). Gluonic excitations of these systems can exist that have different color and parity. However, gluonic radiation does not change the flavor symmetry or isospin of a two-quark system. Hence, we also have the states such as

$$[q,q]_6^- \to (SU_3^{flavor} = \bar{3}, J^P = 1^-) \quad (3.3)$$
$$\{q,q\}_6^- \to (SU_3^{flavor} = 6, J^P = 0^-) \quad (3.4)$$

Whenever we deal with such "excited" diquark systems it is convenient to label both the color representation and parity. During the expansion of a colored system the positive parity and negative parity states and the $\bar{3}_c$ and $6_c$ states of the same flavor symmetry can mix. However, since gluonic interactions do not change the flavor symmetry of diquarks, in the formation of a color singlet baryon, when considering the "fragmentation" of the diquark system into a color-singlet hadron, it is sufficient to consider only the $\bar{3}_c$ component of the combined state. Hence, we can describe baryon production by labeling the diquark content of the target fragmentation region and the flavor of the valence quark that participates in the hard scattering event. For the hard scattering of a valence u quark from the proton we distinguish two fracture functions

$$M^u_{B/[u,d]}(x,z;\mu^2) \, \& \, M^u_{B/\{u,d\}}(x,z;\mu^2)$$

while for the hard scattering of a valence d quark from the proton we have

$$M^{d}_{B/\{u,u\}}(x,z;\mu^2).$$

For a neutron target we distinguish fracture functions

$$M^{u}_{B/\{d,d\}}(x,z;\mu^2), M^{d}_{B/\{u,d\}}(x,z;\mu^2) \ \& \ M^{d}_{B/[u,d]}(x,z;\mu^2).$$

In all cases, the difference between proton and neutron targets can be inferred from the combined flavor content of the scattered quark and remnant diquark. The notation chosen here emphasizes that we are primarily interested in the target fragmentation aspect of the fracture function formalism for this application. For the produced baryons, we are restricting attention to the even parity S-wave states of quark model spectroscopy that can be produced by adding one quark, either a u, d or s quark, to the designated diquark remnant state.

To present the production of polarized baryons in a hard-scattering process in the fracture function formalism we apply the framework developed by Ceccopieri and Trendadue [27] and consider the extension of fracture functions to include transverse momenta in the final state. For convenience, we specialize to the case of semi-inclusive deep inelastic lepton scattering where the orientation of the large space-like momentum transfer is chosen to be along the $\hat{z}$ axis, $\vec{q} = Q\hat{e}_z$. The fracture function, $M^{q}_{B\uparrow/p}(x,\vec{p}_T,z,\vec{k}_T;Q^2)$, as defined in [27] then gives the conjoint probability to find in the hard lepton-scattering process, both a quark jet with longitudinal momentum fraction, x, (along the z-axis) and transverse momentum, $\vec{p}_T$, and a spin-polarized baryon with longitudinal momentum fraction, z, and transverse momentum, $\vec{k}_T$. These two transverse momenta represent independent kinematic observables with $\vec{p}_T$ defined by the transvese component of the thrust axis for the quark jet and $\vec{k}_T$ to be the transverse momentum of the produced baryon in the target fragmentation region with respect to the z axis. For this application, we choose the transverse momentum of the observed baryon to be along the x axis, $\vec{k}_T = k_{TN}\hat{e}_x$. Based on the discussion presented in Sec. II, this choice ensures that production spin density matrix for the polarized baryon can be diagonalized in the $\hat{\Sigma}_y$ basis. We can therefore define the polarizing fracture function in terms of the production spin asymmetry

$$\Delta^N M^{q}_{B\uparrow/p}(x,z,k_{TN};Q^2) = \int d^2\vec{p}_T \{ M^{q}_{B\uparrow/p}(x,\vec{p}_T,z,k_{TN}\hat{e}_x;Q^2) - M^{q}_{B\downarrow/p}(x,\vec{p}_T,z,k_{TN}\hat{e}_x;Q^2) \} \quad (3.5)$$

where the integration over $d^2\vec{p}_T$ averages of the transverse kinematics of the current fragmentation region.

In our application here, we emphasize the relationship of the polarizing fracture functions to diquark fragmentation by specifying, for example,

$$\Delta^N M^u_{n\uparrow/p}(x,z,k_{TN};Q^2) = \Delta^N M^u_{n\uparrow/[u,d]}(x,z,k_{TN};Q^2) + \Delta^N M^u_{n\uparrow/\{u,d\}}(x,z,k_{TN};Q^2), \quad (3.6)$$

$$\Delta^N M^d_{\Delta^{++}/p}(x,z,k_{TN};Q^2) = \Delta^N M^d_{\Delta^{++}/\{u,u\}}(x,z,k_{TN};Q^2), \quad (3.7)$$

and so on for the allowed set of produced baryons. The polarizing fracture functions for timelike momentum transfers, denoted $\Delta^N M^q_{B\uparrow/[q,q]}(x,z,k_{TN};M^2)$ in our nomenclature, require alternate conventions for specifying the orientation of momenta and spins. In future formulas presented here in which the form of the equations does not depend on the symmetry of the diquark system we will use designation (q,q) to denote that the equation is valid for both [q,q] and {q,q} diquarks.

As indicated in Fig.'s 2 and 3, the transverse momenta of particles observed in hard-scattering processes involve a convolution of the intrinsic transverse momentum associated with the initial state of the $SU_3$-color constituents in the target, the range of transverse momentum generated by hard, "semihard" and collinear corrections to the perturbatively-calculable hard-scattering process, and the transverse momentum generated by the color recombination required by hadronization. However, when considering the spin-directed component of momentum transfer that results in the production of a polarized baryon in the target fragmentation region, it is clear that the source of this effect cannot be associated either with the intrinsic transverse momentum in the unpolarized target nucleon or in the kinematics of the hard scattering. This separation results from the observation of Kane, Pumplin and Repko (KPR) [20] that significant transverse spin-directed momenta cannot be generated perturbaively in qcd. From the discussion found in Sec. II, it is inevitable that the source of the spin-directed momentum, $\delta k_{TN}$, must arise from coherent spin-orbit effects directly associated with the formation of the baryon whose spin is measured. We have referred to this property as KPR factorization. The KPR factorization of the $k_{TN}$- dependence for the polarizing fracture functions is distinct from the factorization properties in x,z of ordinary fracture functions as described by Grazzini, Trentadue and Veneziano. [16] The direct extension of the highly successful approach to the Collins functions $\left| Jet(q_i \uparrow) \right\rangle \to \left| Jet^{'}(q_j \downarrow) \right\rangle + \pi_{i,\bar{j}}$ described by Artru, Czyzewski and Yabuki [13] in terms of the original Lund model [14] suggests the ACY model for diquark fragmentation in terms of the mechanism illustrated in Fig. 4. Here a $^3P_0$ $\bar{q}q$ pair is generated in the midst of the color flux connecting the remnant diquark to the current jet. The capture of the produced quark in the rotating $q\bar{q}$ system with $L = \pm 1$, then, necessarily involves a small spin-directed momentum

$$\langle \delta k_{TN} \rangle = 0.20 \pm 0.04 Gev/c \quad (3.8)$$

In general, this is only a small fraction of the transverse momentum of the produced baryon. As indicated in Fig. 5, the combination of intrinsic transverse momentum and the transverse momentum generated during the hard scattering can result in a broad, $Q^2$-dependent momentum distribution for the produced baryon that is shifted for different values of $L_y$. Thus, in the region where the momentum dependence of the production process is steeply falling, the small spin-directed momentum shift $\delta k_{TN}$ can produce significant polarization effects.

Given that the pair production occurs preferentially along the thrust axis of the overall event and
$$\left\langle L_y \right\rangle_{pair} = \left\langle zk_x - xk_z \right\rangle_{pair} \tag{3.9}$$

where the spatial and momentum directions are specified in the CM system of the $q\bar{q}$ pair, the correlation between the spin-directed momentum, $\delta k_{TN}$, and $L_y$ is opposite in the target fragmentation region from that occurring in the current fragmentation region. The content of Fig. 5 includes the quark spin-directed momentum shift of the quark capture processes with $L_y = \pm 1$ compared to the capture processes with $L_y = 0$. In the kinematic regions we consider, the polarizing fracture functions are determined by the flavor and spin of the captured quark combined with the flavor and spin of the remnant diquark system. To focus on the connection between the spin orientation of the produced baryon and the transverse momentum, it is therefore convenient to consider the partial wave decomposition of the fragmentation process

$$M^{q_i}_{B\uparrow/(q_j,q_k)}(x,z,k_{TN};Q^2) = C^{q_i}_{(q_j,q_k)}(x,z;Q^2) G^{(q_j,q_k)}_{B\uparrow}(z,k_{TN};Q^2)$$
$$G^{(q_j,q_k)}_{B\uparrow}(z,k_{TN};Q^2) \cong \sum_{L=-1}^{+1} G^{(q_j,q_k)}_{(L)B\uparrow}(z,k_{TN};Q^2) \tag{3.10}$$

The first line of (3.10) uses KPR factorization to isolate the dependence on $k_{TN}$ into an effective fragmentation function. This KPR factorization can be true even at values of $Q^2$ at which it is not valid to assume factorization in x and z for the fracture functions integrated over transverse momentum. [16] The second line assumes that the quark capture process is saturated by those states in which the $q\bar{q}$ pair is produced with $L_y = 0, \pm 1$. These assumptions capture the basic content of the ACY model for the polarizing fracture functions of baryons.

Taking into consideration the convention connecting the orientation of baryon spin and the $k_{TN}$ applicable in the target fragmentation region, we can then define

$$\delta G^{(q_j,q_k)}_{B\uparrow}(z,k_{TN};Q^2) = G^{(q_j,q_k)}_{(L=-1)B\uparrow}(z,k_{TN};Q^2) - G^{(q_j,q_k)}_{(L=+1)B\uparrow}(z,k_{TN};Q^2)$$
$$= G^{(q_j,q_k)}_{(L=-1)B\uparrow}(z,k_{TN};Q^2) - G^{(q_j,q_k)}_{(L=-1)B\uparrow}(z,-k_{TN};Q^2) \tag{3.11}$$

This then gives the KPR-factorized expression for the polarizing fragmentation functions for the ACY model in the form

$$\Delta^N M^{q_i}_{B\uparrow/(q_j,q_k)}(x,z,k_{TN};Q^2) \cong C^{q_i}_{(q_j,q_k)}(x,z;Q^2)\delta G^{(q_j,q_k)}_{B\uparrow}(z,k_{TN};Q^2) \qquad (3.12)$$

This simple , basic formula demonstrates the practical utility of identifying the origin of the spin-directed momentum in measurements of single-spin observables. A gaussian approximation for $\delta G(k_{TN})$ that corresponds to the curves in Fig. 5 is then shown in Fig. 6. In the next section we will explore the relations between the polarizing fragmentation functions for different baryons implied by the diquark basis for baryon structure combined with the result that the spin density matrices must be diagonal in the transversity frame..

### IV.  Diquark Fragmentation Dynamics and Tests of Baryon Structure.

A thorough and systematic study of the dynamics of baryon production in semi-inclusive deep inelastic processes can provide significant opportunities to explore details of baron structure. The polarizing fracture functions defined in Sec. III are particularly useful in testing the specific hypotheses involved in the quark-diquark approach to baryon dynamics since these involve direct correlations between flavor and spin degrees of freedom. If we assume isospin invariance and broken SU(3) flavor invariance for the strong interactions, there are significant relationships that appear in the fracture functions $M^{q_i}_{B\uparrow/(q_j,q_k)}(x,z,k_{TN};Q^2)$ for baryons in different flavor and spin states. For example, isolating the Bjorken x dependence as in (3.10) leads to relationships such as

$$C^u_{\{u,d\}}(x,z;Q^2) = C^d_{\{u,d\}}(x,z;Q^2) = C^d_{\{u,u\}}(x,z;Q^2) = C^u_{\{d,d\}}(x,z;Q^2) \qquad (4.1)$$

involving the $I=1$ diquark component of the nucleon wave function. In contrast, the additional attraction in the "favored" diquark system leads to the condition

$$\lim_{x\to 1}\left\{C^u_{[u,d]}(x,z;Q^2) \square\; C^u_{\{u,d\}}(x,z;Q^2)\right\} \qquad (4.2)$$

so that different mixtures of diquark quantum numbers can be sampled by varying Bjorken x in the experiment. The flavor and spin quantum numbers of the produced baryons provide additional tests of the underlying assumption that it is possible to separate baryon wave functions in the form

$$|B:JM,I\mathrm{M}\rangle = |(q,q):j_2 m_2, i_2 \mu_2\rangle \otimes |q:j_1 m_1, i_1 \mu_1\rangle \qquad (4.3)$$

reflecting the spin and isospin projections for the quark and diquark systems combining to form the final baryon system. We can use the results of Sec. II that show that the spin density matrix is diagonal when the spin projections are in the transverse, $\hat{y}$, direction to parametrize the spin and isospin dependence of the fragmentation process in a form

$$G_{B\uparrow}^{(q,q)}(z,k_{TN};Q^2) = \left|\langle i_1 i_2; \mu_1 \mu_2 : IM \rangle\right|^2 \left|\langle j_1 j_2; m_1 m_2 : JM \rangle\right|^2 G_S(z,k_{TN};m_B^2,Q^2) \quad (4.4)$$

that directly involves the spin and isospin Clebsch-Gordon coefficients[29] representing the combinations specified. The function, $G_S(z,k_{TN};m_B^2,Q^2)$ can also depend on the strangeness, $S = 0, -1$ of the captured quark and on the $mass^2$ of the produced baryon. For scattering from an unpolarized target, the $J = 1$ diquarks are to be found in an uncorrelated spin density matrix. Combining with a polarized quark then leads to very specific states, such as

$$\left|\{d,d\}_o\right\rangle \otimes \left|u \downarrow\right\rangle = \frac{1}{3}\left|\Delta^o\right\rangle \begin{bmatrix} 0 & 0 & 0 & 0 \\ 0 & 1/6 & 0 & 0 \\ 0 & 0 & 1/3 & 0 \\ 0 & 0 & 0 & 1/2 \end{bmatrix} \oplus \frac{2}{3}\left|n\right\rangle \begin{bmatrix} 2/3 & 0 \\ 0 & 1/3 \end{bmatrix} \quad (4.5)$$

where the elements of the spin density matrix are explicitly displayed. The combination of a $J = 0$ diquark with a polarized quark leads to a simple baryon state such as

$$\left|[u,d]\right\rangle \otimes \left|s \downarrow\right\rangle = \left|\Lambda_o\right\rangle \begin{bmatrix} 0 & 0 \\ 0 & 1 \end{bmatrix} \quad (4.6)$$

These connections lead to very simple expressions for the polarizing fracture functions based on the underlying structure of the ACY model [13]. We shall implement the set of assumptions of this model that the captured quark comes from a $J^P = 0^{+\,3}P_0$ pair with $L_y = +1(q \downarrow), L_y = -1(q \uparrow)$ or from a mixture of states with $L_y = 0(q_{unpolarized})$. The corresponding spin density matrices occurring for the production of a $J = \frac{3}{2}(\Delta, \Sigma^*)$ baryon are then

$$\rho^{3/2}_{\{\}}(L=+1) = diag\begin{bmatrix}0\\1/6\\1/3\\1/2\end{bmatrix} \quad \rho^{3/2}_{\{\}}(L=0) = diag\begin{bmatrix}1/4\\1/4\\1/4\\1/4\end{bmatrix}$$

$$\rho^{3/2}_{\{\}}(L=-1) = diag\begin{bmatrix}1/2\\1/3\\1/6\\0\end{bmatrix} \quad \delta\rho^{3/2}_{\{\}} = diag\begin{bmatrix}1/2\\1/6\\-1/6\\-1/2\end{bmatrix}$$

(4.7)

where diag denotes the diagonal elements of the 4x4 matrices with all off-diagonal entries set to 0. For the production of a $J = \frac{1}{2}(p,n,\Sigma)$ from combining with a $\{q,q\}$ J=1 diquark we get

$$\rho^{1/2}_{\{\}}(L=+1) = diag\begin{bmatrix}2/3\\1/3\end{bmatrix} \quad \rho^{1/2}_{\{\}}(L=0) = diag\begin{bmatrix}1/2\\1/2\end{bmatrix}$$

$$\rho^{1/2}_{\{\}}(L=-1) = diag\begin{bmatrix}1/3\\2/3\end{bmatrix} \quad \delta\rho^{1/2}_{\{\}} = diag\begin{bmatrix}-1/3\\1/3\end{bmatrix}$$

(4.8)

The production of a $J = \frac{1}{2}(p,n,\Lambda)$ baryon from a [u,d], $J=0$ diquark then leads to the density matrices

$$\rho^{1/2}_{[]}(L=+1) = diag\begin{bmatrix}0\\1\end{bmatrix} \quad \rho^{1/2}_{[]}(L=0) = diag\begin{bmatrix}1/2\\1/2\end{bmatrix}$$

$$\rho^{1/2}_{[]}(L=-1) = diag\begin{bmatrix}1\\0\end{bmatrix} \quad \delta\rho^{1/2}_{[]} = diag\begin{bmatrix}1\\-1\end{bmatrix}$$

(4.9)

Using these expressions we can write the KPR factorized form for the ACY model for the polarizing facture functions using the matrices, $\delta\rho^{3/2}_{\{\}}$, $\delta\rho^{1/2}_{\{\}}$, and $\delta\rho^{1/2}_{[]}$. In these equations we will suppress the dependence on x,z and $Q^2$ as shown in (3.12) and (4.4) and only display the dependence of the functions on $k_{TN}$. For the production of polarized $J = \frac{3}{2}$ baryons from an unpolarized proton target we get,

$$\Delta^N M^d_{\Delta^{++}\uparrow/\{u,u\}} = C^d_{\{u,u\}}\delta G_0(k_{TN})\delta\rho^{3/2}_{\{\}}$$

$$\Delta^N M^d_{\Delta^{+}\uparrow/\{u,u\}} = \frac{1}{3}C^d_{\{u,u\}}\delta G_0(k_{TN})\delta\rho^{3/2}_{\{\}}$$

(4.10)

$$\Delta^N M^d_{\Sigma^{*+}/\{u,u\}} = C^d_{\{u,u\}}\delta G_{-1}(k_{TN})\delta\rho^{3/2}_{\{\}}$$

In these expressions $\delta G_0(k_{TN})$ describes the $k_{TN}$ dependence resulting from the preferential capture of a nonstrange u or d quark in the ACY model by (3.11) and

$\delta G_{-1}(k_{TN})$ describes that result from the capture of an s quark with strangeness $-1$. In general, these will be different.

$$\Delta^N M^u_{\Delta^+\uparrow/\{u,d\}} = \frac{2}{3} C^u_{\{u,d\}} \delta G_0(k_{TN}) \delta\rho^{3/2}_{\{\}}$$

$$\Delta^N M^u_{\Delta^o\uparrow/\{u,d\}} = \frac{2}{3} C^u_{\{u,d\}} \delta G_0(k_{TN}) \delta\rho^{3/2}_{\{\}} \tag{4.11}$$

$$\Delta^N M^u_{\Sigma^{*o}/\{u,d\}} = C^u_{\{u,d\}} \delta G_{-1}(k_{TN}) \delta\rho^{3/2}_{\{\}}$$

The corresponding set of expressions for the polarizing fracture functions of decuplet baryons from a neutron target are

$$\Delta^N M^d_{\Delta^+\uparrow/\{u,d\}} = \frac{2}{3} C^d_{\{u,d\}} \delta G_0(k_{TN}) \delta\rho^{3/2}_{\{\}}$$

$$\Delta^N M^d_{\Delta^o\uparrow/\{u,d\}} = \frac{2}{3} C^d_{\{u,d\}} \delta G_0(k_{TN}) \delta\rho^{3/2}_{\{\}} \tag{4.12}$$

$$\Delta^N M^d_{\Sigma^{*o}/\{u,d\}} = C^d_{\{u,d\}} \delta G_{-1}(k_{TN}) \delta\rho^{3/2}_{\{\}}$$

and

$$\Delta^N M^u_{\Delta^o\uparrow/\{d,d\}} = \frac{1}{3} C^u_{\{d,d\}} \delta G_0(k_{TN}) \delta\rho^{3/2}_{\{\}}$$

$$\Delta^N M^u_{\Delta^-\uparrow/\{d,d\}} = C^u_{\{d,d\}} \delta G_0(k_{TN}) \delta\rho^{3/2}_{\{\}} \tag{4.13}$$

$$\Delta^N M^u_{\Sigma^{*-}/\{d,d\}} = C^u_{\{d,d\}} \delta G_{-1}(k_{TN}) \delta\rho^{3/2}_{\{\}}$$

All these expressions involve the matrix $\delta\rho^{3/2}_{\{\}}$ shown in (4.7). Thus the polarizing fracture functions for these decuplet baryons provide a set of specific tests regarding the isospin constraints built into the model.

Turning to the polarizing fragmentation functions for the production of octet baryons we get, for the production of polarized protons from an unpolarized proton target, the expressions

$$\Delta^N M^d_{p\uparrow/\{u,u\}} = \frac{2}{3} C^d_{\{u,u\}} \delta G_0(k_{TN}) \delta\rho^{1/2}_{\{\}}$$

$$\Delta^N M^u_{p\uparrow/\{u,d\}} = \frac{1}{3} C^u_{\{u,d\}} \delta G_0(k_{TN}) \delta\rho^{1/2}_{\{\}} \tag{4.14}$$

$$\Delta^N M^u_{p\uparrow/[u,d]} = C^u_{[u,d]} \delta G_0(k_{TN}) \delta\rho^{1/2}_{[]}$$

These equations involve the matrices $\delta\rho^{1/2}_{\{\}}$ from (4.8) and $\delta\rho^{1/2}_{[]}$ from (4.9). Since these are opposite in sign, there is an interplay from the different Bjorken x dependence of $C^u_{\{u,d\}}(x,z;Q^2)$ and $C^u_{[u,d]}(x,z;Q^2)$ that reflects stronger binding of the favored $[u,d]$ diquark. The polarizing fracture functions for polarized neutrons from a proton target are

$$\Delta^N M^u_{n\uparrow/\{u,d\}} = \frac{1}{3} C^u_{\{u,d\}} \delta G_0(k_{TN}) \delta\rho^{1/2}_{\{\}}$$
$$\Delta^N M^u_{n\uparrow/[u,d]} = C^u_{[u,d]} \delta G_0(k_{TN}) \delta\rho^{1/2}_{[]}$$
(4.15)

The fracture functions for polarized $\Lambda, \Sigma$ hyperons from a proton target are then

$$\Delta^N M^u_{\Lambda\uparrow/[u,d]} = C^u_{[u,d]} \delta G_{-1}(k_{TN}) \delta\rho^{1/2}_{[]}$$
$$\Delta^N M^d_{\Sigma^+\uparrow/\{u,u\}} = C^d_{\{u,u\}} \delta G_{-1}(k_{TN}) \delta\rho^{1/2}_{\{\}}$$
$$\Delta^N M^u_{\Sigma^0\uparrow/\{u,d\}} = C^u_{\{u,d\}} \delta G_{-1}(k_{TN}) \delta\rho^{1/2}_{\{\}}$$
(4.16)

The opposite sign for the polarization of Lambdas and Sigmas found in the Lund model[14] is displayed directly by inserting the expressions (4.8) and (4.9) for the matrices $\delta\rho^{1/2}_{\{\}}$ and $\delta\rho^{1/2}_{[]}$ into these equations. Note that the different Bjorken x dependence of $C^u_{[u,d]}(x,z;Q^2)$ compared to $C^u_{\{u,d\}}(x,z;Q^2)$ predicted in (4.2) reflects the binding of the favored [u,d] diquark and is implied in (4.16).

The analogous expressions involving neutron targets are easily written. For the production of polarized neutrons from a neutron target we have,

$$\Delta^N M^u_{n\uparrow/\{d,d\}} = \frac{2}{3} C^u_{\{d,d\}} \delta G_0(k_{TN}) \delta\rho^{1/2}_{\{\}}$$
$$\Delta^N M^d_{n\uparrow/\{u,d\}} = \frac{1}{3} C^d_{\{u,d\}} \delta G_0(k_{TN}) \delta\rho^{1/2}_{\{\}}$$
$$\Delta^N M^d_{n\uparrow/[u,d]} = C^d_{[u,d]} \delta G_0(k_{TN}) \delta\rho^{1/2}_{[]}$$
(4.17)

The expressions for the production of polarized protons from a neutron target are given as

$$\Delta^N M^d_{p\uparrow/\{u,d\}} = \frac{1}{3} C^d_{\{u,d\}} \delta G_0(k_{TN}) \delta\rho^{1/2}_{\{\}}$$
$$\Delta^N M^d_{p\uparrow/[u,d]} = C^d_{[u,d]} \delta G_0(k_{TN}) \delta\rho^{1/2}_{[]}$$
(4.18)

For polarized hyperons from a neutron target we have the expressions

$$\Delta^N M^d_{\Lambda\uparrow/[u,d]} = C^d_{[u,d]} \delta G_{-1}(k_{TN}) \delta\rho^{1/2}_{[]}$$
$$\Delta^N M^u_{\Sigma^-\uparrow/\{d,d\}} = C^u_{\{d,d\}} \delta G_{-1}(k_{TN}) \delta\rho^{1/2}_{\{\}}$$
$$\Delta^N M^d_{\Sigma^0\uparrow/\{u,d\}} = C^d_{\{u,d\}} \delta G_{-1}(k_{TN}) \delta\rho^{1/2}_{\{\}}$$
(4.19)

The 28 different expressions for the polarizing fracture functions in the ACY model displayed in Eqs. (4.10)- (4.19) constitute the main content of this paper. These expressions embody the power of KPR factorization and the diagonalization of the spin density matrix into a tightly constrained set of formulae representing numerous distinguishable measurements. They directly illustrate the value of KPR factorization and provide an explicit example of an application of the concept of spin-directed momentum.

These polarizing fracture functions represent direct observables. The weak decays of $\Lambda, \Sigma$ provide measurements of their polarizations. The spin orientation of decuplet baryons in deep inelastic processes can be inferred from a partial wave analysis of known final states in the strong decays. Proton and neutron spin orientation measurements require rescattering from a material with known analyzing power. The Bjorken-x and Feynman-z dependence of these production processes contains significant information about chiral mechanisms mentioned in Sec. I. For discussion purposes, it is convenient to display the full $x, z$ dependence of the expressions for the production of $\Lambda\uparrow, \Sigma\uparrow$ from a proton target.

$$\Delta^N M^u_{\Lambda\uparrow/[u,d]}(x,z,k_{TN};Q^2) = C^u_{[u,d]}(x,z;Q^2)\delta G_{-1}(z,k_{TN};Q^2)diag\begin{bmatrix}1\\-1\end{bmatrix}$$

$$\Delta^N M^u_{\Sigma\uparrow/\{u,d\}}(x,z,k_{TN};Q^2) = C^u_{\{u,d\}}(x,z;Q^2)\delta G_{-1}(z,k_{TN};Q^2)diag\begin{bmatrix}-1/3\\1/3\end{bmatrix}$$

(4.20)

We now turn to a brief discussion of these mechanisms.

### Sec. V.  Spin Observables and Chiral Mechanisms in QCD.

The polarizing fracture functions for baryon production in deep inelastic processes provide an approach to studying baryon structure and the consequences of chiral symmetry in qcd. The existence of a symmetry in quantum field theory always simplifies the analysis of dynamical mechanisms and frequently allows calculations that yield approximate results even when the theory cannot be solved exactly. The implications of the broken chiral symmetry in qcd have been extensively studied [29] in many ways. Much of our understanding of the nonperturbative aspects of the strong interactions is obtained from models embodying the techniques developed in chiral models. When applying these studies to the internal structure of hadrons and hadronic systems, it is worth noting that strong interaction dynamics contains many examples of large mass differences between systems with similar quantum numbers. The most familiar examples are,

$$M_\rho - M_\pi \cong 639 MeV, \quad (5.1)$$

the mass difference between vector and pseudoscalar mesons and the mass difference between a "constituent" quark $Q = U, D$ and a "current" or "partonic" quark $q = u, d$

$$M_Q - M_q \cong 320 MeV \quad (5.2)$$

Jaffe and Wilczek [24,25,26] have recently emphasized that diquarks must also be considered as contributing to the internal chiral structure of hadrons. The evidence for this hypothesis includes the difference in the effective mass for diquarks with differing spin structure,

$$M_{\{u,d\}} - M_{[u,d]} \cong M_{\Delta^+} - M_p \cong 294 MeV$$
$$M_{\{s,u\}} - M_{[s,u]} \cong M_{\Sigma^{*+}} - M_{\Sigma^+} \cong 196 MeV$$
(5.3)

These effective diquark masses lead to the different predictions for the Bjorken x dependence of $C^u_{\{u,d\}}(x,z;Q^2)$ and $C^y_{[u,d]}(x,z;Q^2)$ as given in Eq. (4.1). The study of baryon polarization in the target fragmentation of deep-inelastic lepton scattering provides, in principle, more experimental control than can be achieved in inclusive hadronic processes. For example, the polarizing fracture functions for $\Lambda\uparrow,\Sigma\uparrow$ as displayed in (4.20) should display both the opposite spin polarization and the difference in Bjorken x-dependence associated with the mass difference (5.3). The large-x behavior of the polarizing fracture functions for the decuplet baryons should all follow that of $\Delta^N M^u_{\Sigma\uparrow/\{u,d\}}(x,z,k_{TN};Q^2)$ while displaying polarizations of opposite sign. The competition of the different sets of polarizing fracture functions for $p\uparrow$ and $n\uparrow$ from a proton target as given in (4.14) and (4.15) can be explored by the difference in the predicted x dependence. The spin dependence of baryon production provides multiple tests of dipole structure in a manner complementary to the study of high-mass baryon spectroscopy.

It is notable that the complex landscape of the phenomenology involving the approximate chiral symmetry of qcd reflects the fact that many hadronic states play more than one "role" in the interplay of theory and experiment. The simplest example of the "multi-role" player involves the interactions of pions. The pion plays the role of pseudo-Goldstone boson of the broken symmetry [30] in the low momentum interaction of hadrons but displays its $q\bar{q}$ character in many other processes. A more complicated example involves the $a_o/f_o(600)$ state that appears as a broad enhancement in the $\pi\pi$ channel in many processes. This state can be associated with the $\sigma$ meson of various versions of the $\sigma$ model [31]. It has also been considered a possible $\pi-\pi$ "hadronic molecule", a gluonic enhancement or an instanton-generated multiquark state. [32] Recent attention has also been directed at its possible role as a $[u,d]-[\bar{u},\bar{d}]$ "tetraquark" state [25].

As shown above, this low-mass enhancement can also play an important role in the study of single-spin observables. It appears in the ACY model for Collins Functions [13], for the polarizing fragmentation functions for vector mesons [33] as well as in the construction of polarizing fracture functions for baryons presented here. In all the spin observables, a low-mass $^3P_0$ $q\bar{q}$ state with $J^{PC}=0^{++}$ is required to explain the relative normalization of the spin-directed momentum appearing in related processes. Our understanding of nonperturbative qcd must therefore accommodate the possibility that all the possible channels connecting to these quantum numbers interact strongly and are strongly mixed.

The dynamics that appear directly in the polarizing fracture functions also appear as virtual corrections to the wave function of the proton [19] leading to constituent orbital angular momentum and the orbital structure functions and Boer-Mulders functions. The study included in this paper represents a necessary step in completing the program outlined there. However, it is only a preliminary step. As indicated in the introduction, this work has the limited goal of illustrating the concept of spin-directed momentum in a specific framework. A more systematic and phenomenological approach to Collins functions, polarizing fragmentation functions and polarizing fracture functions is found in [33].

**Acknowledgement**

The author has benefited from the considerable insight of G. Bunce in numerous discussions on this subject. Of course, he helped start the whole thing.

$$\Delta^N D_{h/q\uparrow}(z, k_{TN}) = \frac{2k_{TN}}{zM_h} H_1^{Tq}(z, k_{TN}) \qquad \Delta^N D_{h\uparrow/q}(z, k_{TN}) = \frac{k_{TN}}{2M_h} D_1^{Tq}(z, k_{TN})$$

*Collins Functions*          *Polarizing Fragmentation Functions*

$$\Delta^N G_{q\uparrow/p}(x, k_{TN}) = -\frac{k_{TN}}{M_p} h_1^{Tq}(x, k_{TN}) \qquad \Delta^N G_{q/p\uparrow}(x, k_{TN}) = -\frac{k_{TN}}{2M_p} f_1^{Tq}(x, k_{TN})$$

*Boer–Mulders Functions*          *Orbital Structure Functions*

"chiral-odd" functions          "chiral-even" functions

Table 1, The Mulders-Tangerman Quartet

This table gives the relationship between the expressions used by the author for partonic number densities in terms of the expressions commonly used for the related correlators that have the dimension of momentum. The conventions for the signs and the factors of 2 in this table are explained in the article by A. Bacchetta, U. D'Alesio, M. Diehl and C.A. Miller, Phys. Rev. **D70**, 117504 (2004) that defines the "Trento Conventions". The conventions for the polarizing fracture functions defined in this paper are intended to be compatible with the Trento conventions.

|         | T | C | P | CPT | O | $A_\tau$ | $A_\supset$ | $A_\subset$ |
|---------|---|---|---|-----|---|----------|-------------|-------------|
| $\Sigma_x$ | − | + | − | + | − | + | − | + |
| $\Sigma_y$ | + | + | + | + | − | − | − | − |
| $\Sigma_z$ | + | − | − | + | − | + | + | − |

Table II. Finite Symmetries

    For a process involving the measurement of a single spin, rotational invariance and finite symmetries provide strong constraints on the behavior of the components of the measured spin. These are shown for a process occurring in the x-z plane in the table above. We here supplement the usual space-time symmetries of T "time-reflection", C "charge conjugation" and P, "parity" by including symmetries formed by using the Hodge dual operator as explained in Sec. II. The behavior of the spin-reflection operator O is explained more fully there. The operator $A_\tau$ defines a spin-directed momentum for transverse spin observables. The operator $A_\supset = A_\tau T$ and $A_\subset = A_\tau C$ can also be used for defining projection operators in spin calculations.

Figure Captions.

Fig. 1
This sketch displays explicitly how spin-orbit correlations involving constituents of a larger system produce a spin-directed momentum. In fragmentation functions, single-spin asymmetries arise when $\vec{L}$ occurs in the final state of the fragmentation process. Internal orbital angular momentum of colored constituents in the proton also leads to single-spin observables.

Fig. 2
A drawing that indicates the hard-scattering diagram for the production of a polarized baryon, shown as a $\Lambda\uparrow$, in the target fragmentation region of deep-inelastic scattering. We are interested in the transverse momentum of the produced $\Lambda\uparrow$. One contribution to the transverse momentum comes from the intrinsic transverse momentum of the remnant diquark, indicated as region A in the diagram. If we choose the z axis of the production process to be along $\vec{Q}$, another contribution is generated when additional particles are produced in the hard-scattering event in region B of the figure. Finally, a contribution to the transverse momentum of $\Lambda\uparrow$ occurs during the color rearrangement that occurs in the formation of a color-singlet system as indicated in region C of the drawing. A nonvanishing spin-directed momentum transfer, $\delta k_{TN}$, can <u>only</u> occur from coherent spin-orbit dynamics in region C. Although the separation of these regions is not unique, the knowledge that $\delta k_{TN}$ originates in connection with the $\Lambda\uparrow$ in region C provides a powerful constraint on theoretical calculations.

Fig. 3
This sketch supplements the drawing in Fig. 2 by showing a sample separation into regions A,B,C for the components of transverse momentum flow involving particles in the final state of a deep inelastic event.

Fig. 4

The basic assumptions in the extension of the Artru, Czyzewski, Yabuki (ACY) model to polarized fracture functions are indicated in this sketch. A partial wave expansion of the capture of a quark by the target diquark is performed in the CM of the $q\bar{q}$ pair occurring in the color flux produced during the scattering process. Mechanisms involving a $^3P_o$ $q\bar{q}$ pair can dominate the $L_y = \pm 1$ channels. The spin of the quark produced by these mechanisms is opposite to the direction of $L_y$ in these channels. The contributions to $L_y = 0$ channels are assumed to produce an unpolarized ensemble of quarks. The resulting production spin density matrices are diagonal in $\Sigma_y$ of the produced baryon.

Fig. 5

A gaussian approximation to the $k_{TN}$ dependence functions $G_{(L)}$ found in (3.10). The width of the gaussian is one and the L-dependent shift $\delta k_{TN}$ is chosen as 0.2. The relative normalization of the $L = \pm 1$ curves and $L = 0$ is arbitrary. The resulting shape of $\delta G(k_{TN})$ defined in (3.11) is shown in Fig. 6.

Fig. 6

The shape of $\delta G(k_{TN})$ resulting from the curves in Fig. 5.

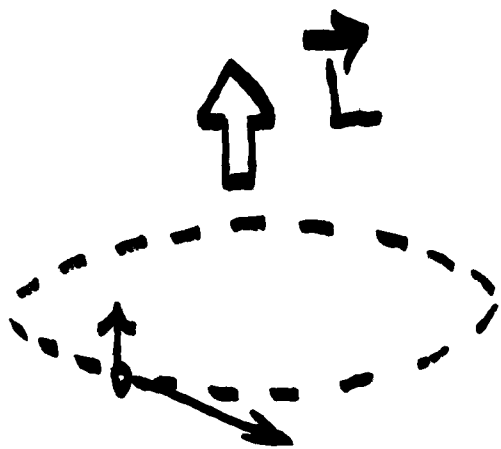 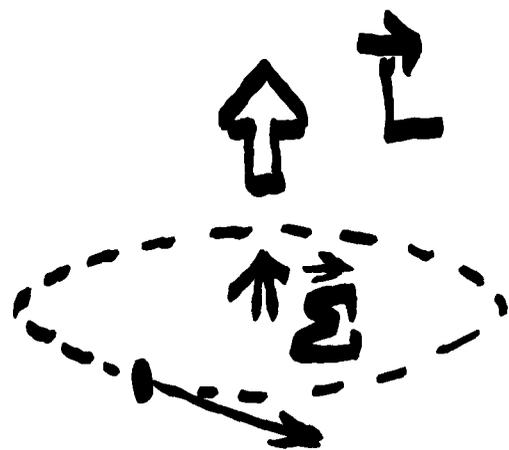

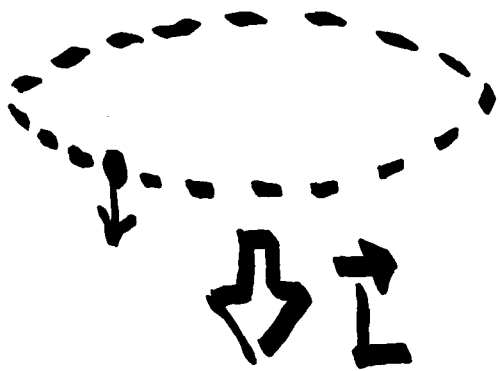 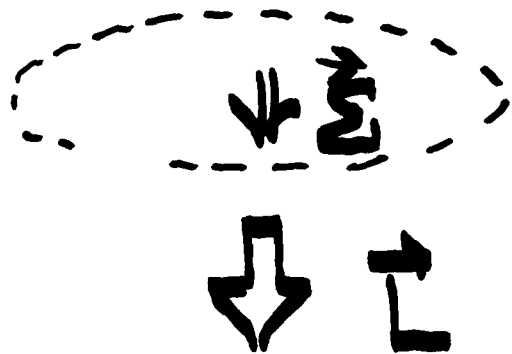

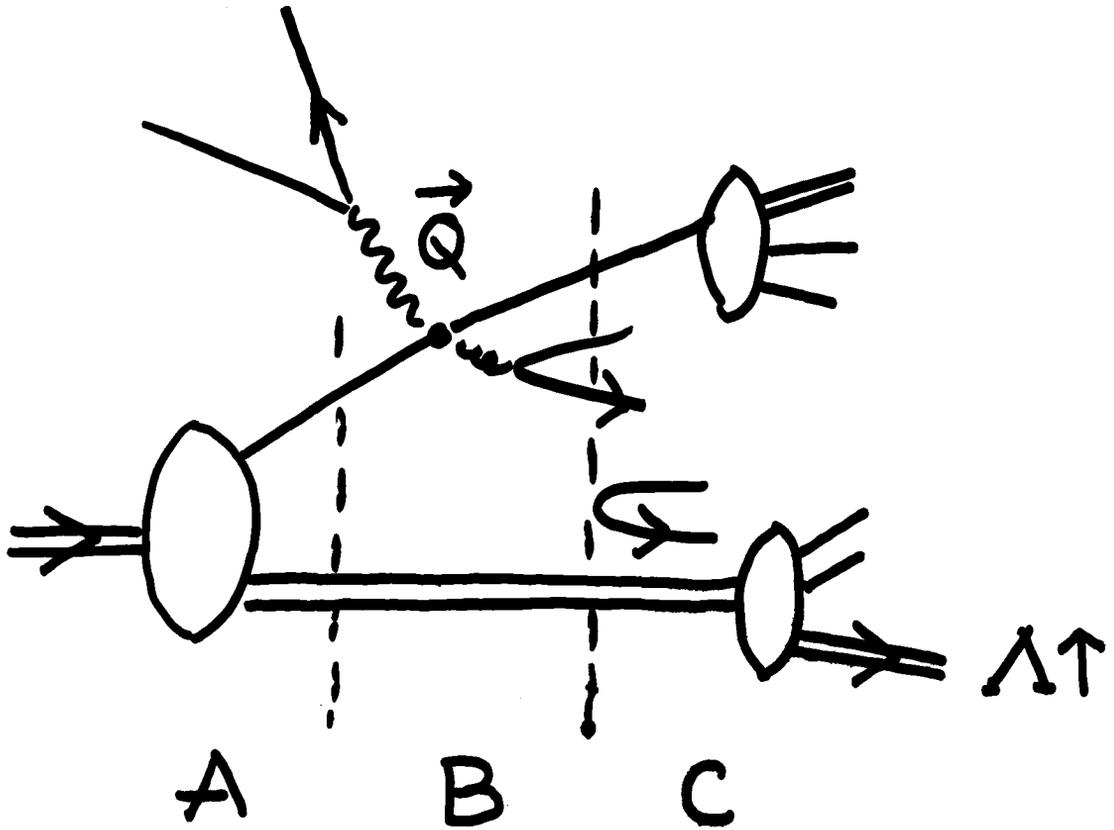

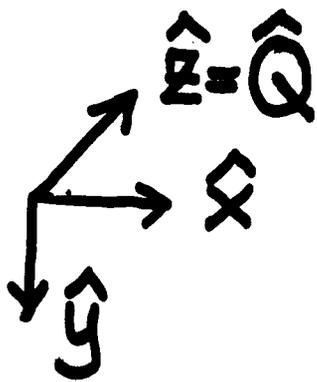
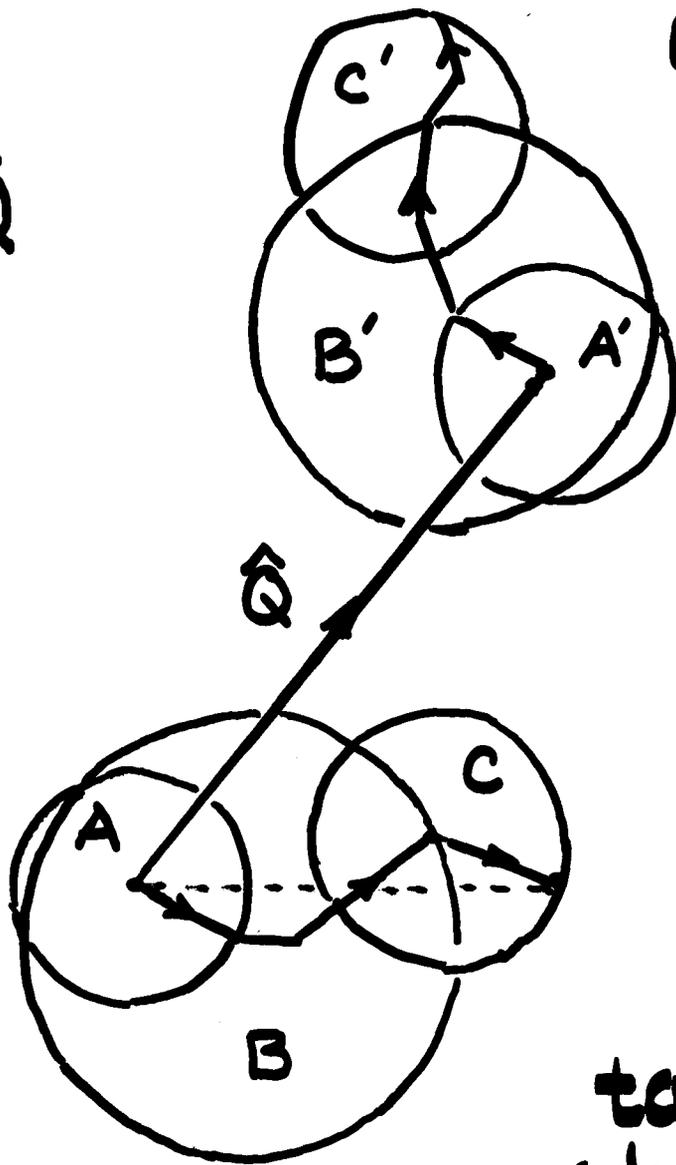

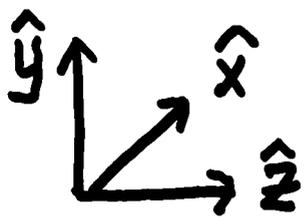

## L=+1

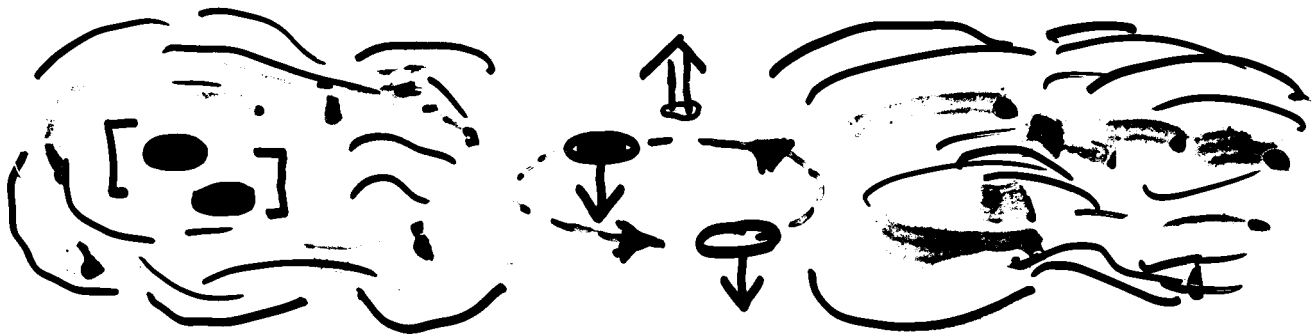

## L=0

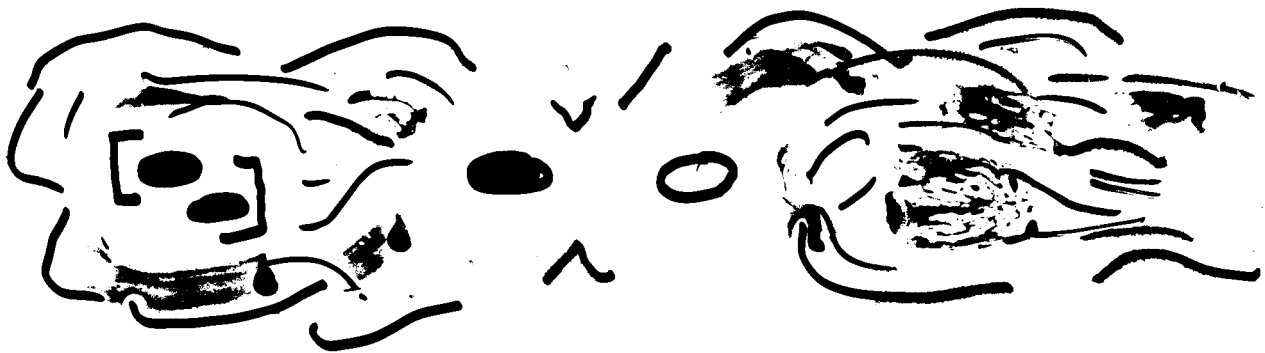

## L=−1

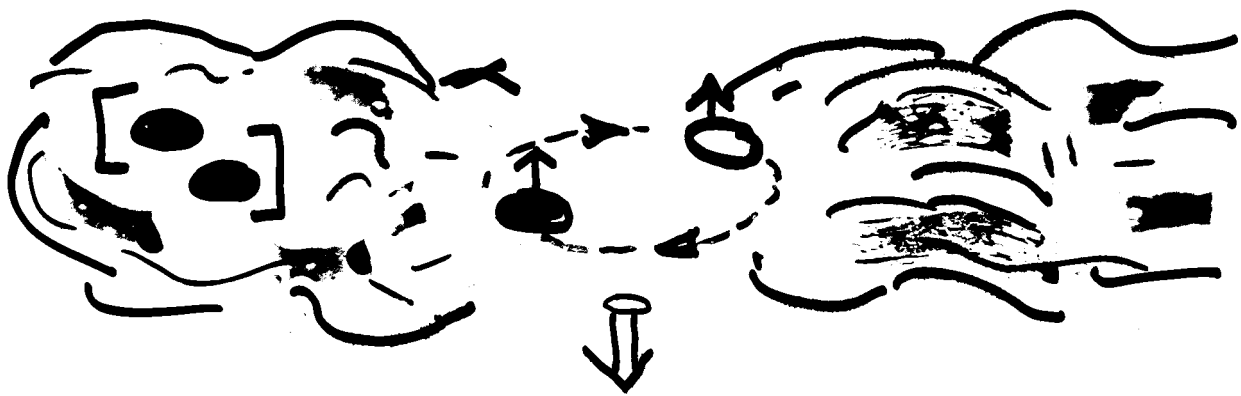

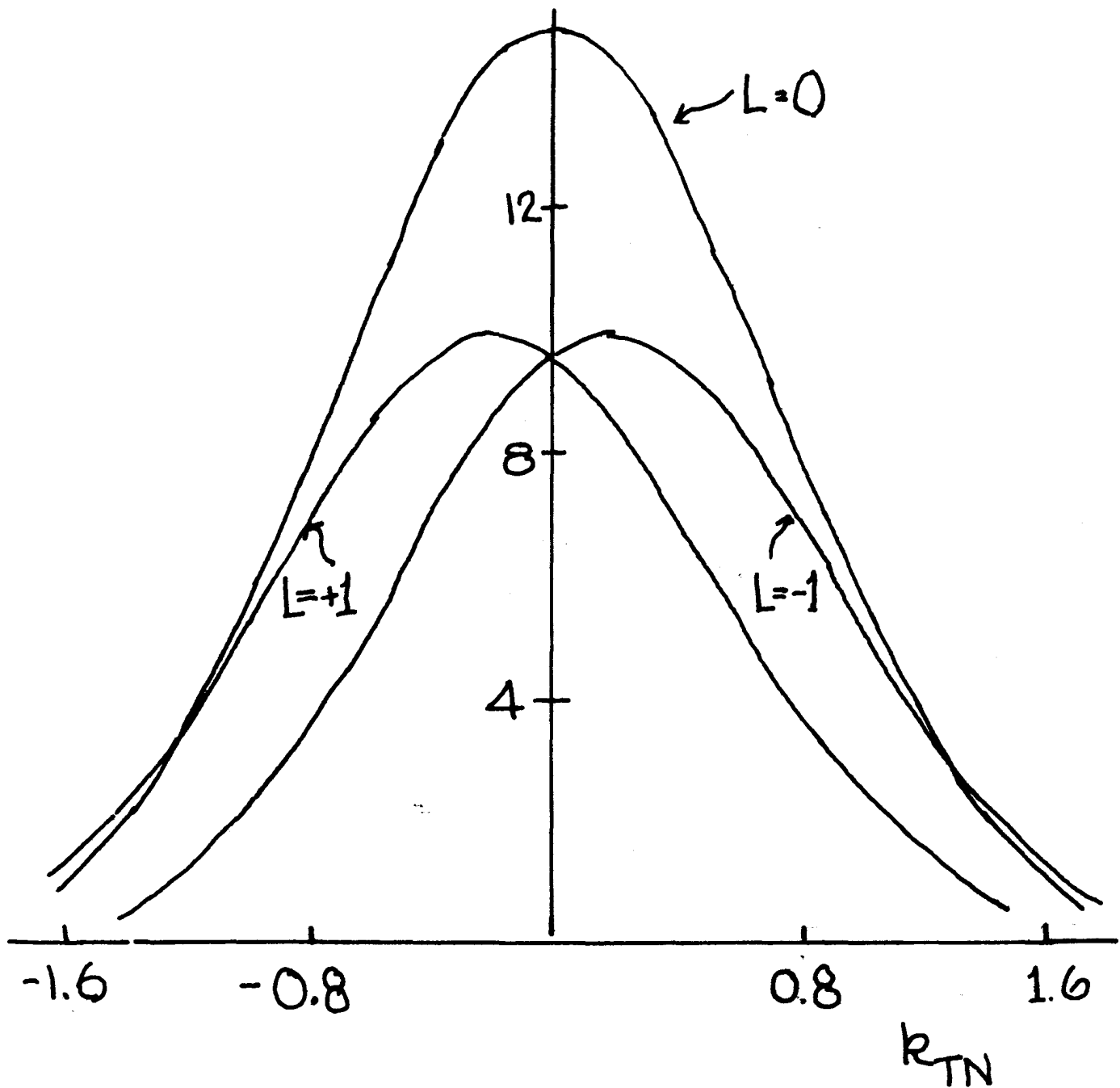

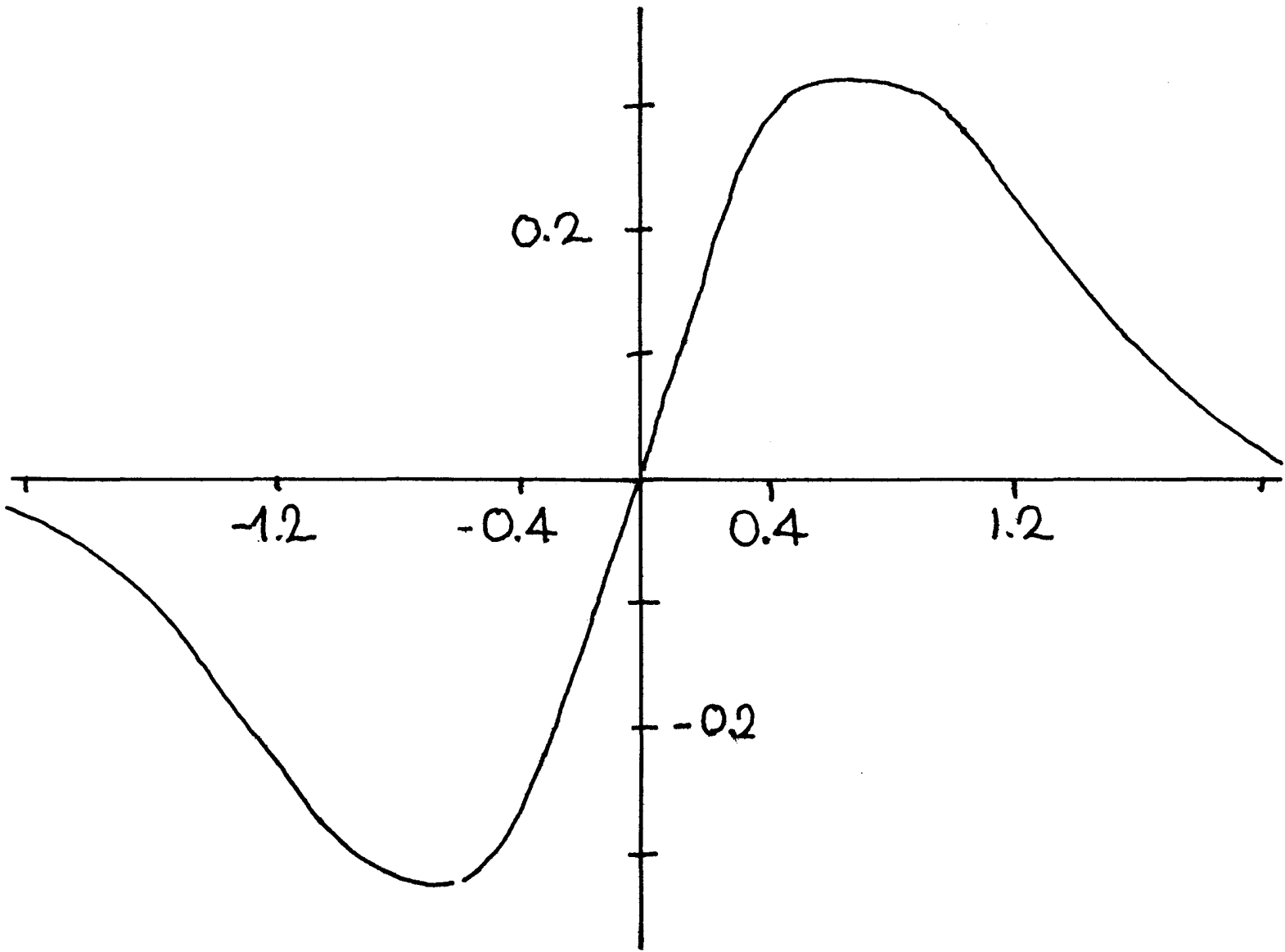